\UseRawInputEncoding
\documentclass[proof]{pasj01}
\draft 
\Received{$\langle$reception date$\rangle$}
\Accepted{$\langle$acception date$\rangle$}
\Published{$\langle$publication date$\rangle$}
\SetRunningHead{Y. Moritani et al.}{near-infrared constraints on stellar wind parameters}

\usepackage{lineno}
\usepackage[normalem]{ulem}
\newcommand{\onefgl}{1FGL~J1018.6$-$5856}

\begin{document}

\title{Novel application to estimate the mass-loss and the dust-formation rates in O-type gamma-ray binaries using near-infrared photometry}
\author{Yuki Moritani$^{1, 2}$, 
Akiko Kawachi$^{3}$, Atsuo T. Okazaki$^{4}$, 
Sho Chimasu$^{3}$ and Hiromi Yoshida$^{3}$ 
}%
\altaffiltext{1}{Subaru Telescope, National Astronomical Observatory of Japan, 650 North A'ohoku Place
Hilo, HI 96720, U.S.A.}
\altaffiltext{2}{Hiroshima Astrophysical Science Center, Hiroshima University, 1-3-1 Kagamiyama Higashi-Hiroshima, Hiroshima 739-8526, Japan}
\altaffiltext{3}{Department of Physics, Tokai University, Hiratsuka, Kanagawa 259-1292, Japan}
\altaffiltext{4}{Faculty of Engineering, Hokkai-Gakuen University, Toyohira-ku, Sapporo 062-8605, Japan}
\email{moritani@naoj.org}


\maketitle

\begin{abstract}
We have performed the near-infrared photometric monitoring observations of two TeV gamma-ray binaries with O-stars (LS~5039 and 1FGL~J1018.6$-$5856), using IRSF/SIRIUS at SAAO, in order to study the stellar parameters and their perturbations caused by the binary interactions.
The whole orbital phase was observed multiple times and no significant variabilities including orbital modulations are detected for both targets. 
Assuming that the two systems are colliding wind binaries, we estimate the amplitude of flux variation caused by the difference in the optical depth of O-star wind at inferior conjunction, where the star is seen through the cavity created by pulsar wind, and other orbital phases without pulsar-wind intervention. The derived amplitude is $<$ 0.001 mag, which is about two orders of magnitude smaller than the observed upper limit. Also using the upper limits of the near-infrared variability, we for the first time obtain the upper limit of the dust formation rate resulting from wind-wind collision in O-star gamma-ray binaries. 
\end{abstract}


\section{Introduction}
Gamma-ray binaries are the group of binaries emitting high-energy (0.1--100~GeV) or/and very high-energy (\textgreater 100~GeV) gamma-rays \citep{Dubus2013,Paredes2019}. About 10 systems have been classified as gamma-ray binaries and all of them are high-mass X-ray binaries (HMXBs). 
The optical-infrared companion stars are either an O star or a Be star---a B-type star exhibiting emission-line profiles in Balmer lines from its dense circumstellar disk (Be disk). 
In a gamma-ray binary, the relativistic particles from the compact object interact 
with the photon field and the low-density, fast outflow (stellar wind) of the massive companion. 
In a Be/gamma-ray binary, they also interact with the high-density, decretion disk around the Be star.
In a system where the compact object is a rotational-powered pulsar, when the pulsar wind collides with the stellar wind, particle acceleration is thought to occur at the shock region resulting in an increase of the non-thermal emission (pulsar wind model).
On the other hand, in a system where mass is transferred from the stellar wind and/or the Be disk to a black hole (or accreting neutron star), sufficient mass transfer leads to jet formation which causes particle acceleration (microquasar model). 
 
Regardless of the nature of the compact object and hence the system, the stellar wind plays an important role in the production of high-energy emission.
In colliding wind binaries, the momentum rate ratio between the pulsar wind and the stellar wind is one of the key parameters for binary nature and geometry of the colliding wind region (e.g. \cite{Dubus2013}). 
Strong shocks, turbulent mixing and secondary shocks in the turbulent flow are developed in this extended asymmetric region \citep{Bosch-Ramon2012,Huber2021,Chen2021}.
In microquasars, \citet{Bosch-Ramon2016} discussed a few impacts of the stellar wind, such as 
on recollimation, bending of the jet, or enhancement of the Coriolis effect by the wind ram pressure, although the mass-loss rate and the velocity of massive star outflows are among the major source of uncertainty.

The stellar wind parameters have been studied using line profiles in the UV and optical bands, but a long-term precise monitoring observation is very difficult to carry out for spectroscopy either in the UV, optical, or near-infrared band, so the relationship between the stellar wind and high-energy emission is still to be discussed in detail.

Both LS~5039 ($e \sim 0.35$, $P_{\mathrm{orb}} =3.9$ days) \citep{Casares2005} and 1FGL~J1018.6$-$5856 ($e = 0.531$, $P_{\mathrm{orb}} = 16.5507$ days) \citep{vanSoelen2022} are gamma-ray binaries hosting an O-star at low Galactic latitude, and their orbit is moderately eccentric.
LS~5039 is a run away star \citep{Ribo2002} and its distance is estimated as 2.0 kpc using parallax of Gaia DR2 and early DR3, which is consistent with the distance estimated from extinction \citep{McSwain2004}.
1FGL~J1018.6$-$5856 is a moving away star at the distance of $6.4^{+1.7}_{-0.7}$ kpc \citep{Marcote2018}. 
\citet{Marcote2018} studied the proper motion of 1FGL~J1018.6$-$5856 precisely using UCSC4, LBA and Gaia DR2 and found that the source has no relation with the supernova remnant in the same line of sight (SNR~G284.3-1.8).

The nature of the compact object has been discussed for these systems.
For instance, \citet{Casares2005} proposed that the compact object in LS~5039 is a black hole (BH), which would place the system in the microquasar class, whereas some argue that the system is a colliding-wind binary with a non-accreting pulsar (e.g., \cite{Martocchia2005, Dubus2006, Sierpowska-Bartosik2007}).
The recent \textit{Suzaku} and \textit{NuSTAR} observations suggested a possible detection of $\sim$~9s X-ray periodicity associating it with a rotating neutron star (NS) \citep{Yoneda2020}, but \citet{Volkov2021} found that the statistical significance of this periodicity is too low.
The orbital solution of 1FGL~J1018.6$-$5856 from the radial velocity of the O star favours a NS but accepts a BH as the nature of the compact object for low inclination angles \citep{Monageng2017}.

Non-thermal emissions from these two systems in radio through gamma-ray bands modulate in the orbital period.
The very high energy (VHE) gamma-ray luminosity from LS~5039 modulates between $(4-10) \times 10^{33}\,\mathrm{erg\,s}^{-1}$ \citep{Aharonian2006}, with a peak around phase 0.8 \citep{Aharonian2006, Casares2005}, where phase 0 corresponds to the periastron passage.
X-ray light curve also shows a peak around similar phase ($\sim0.8$) \citep{Bosch-Ramon2005, Takahashi2009}, while the flux peak in the GeV band takes place around phase 0 \citep{Abdo2009}.
The system shows persistent radio emission, the morphology of which is consistent with mildly relativistic jets \citep{Paredes2000, Paredes2002}.
On the other hand, in 1FGL~J1018.6$-$5856, the GeV and TeV emissions are the highest around phase 0.1 and 0.9, respectively.
In X-rays, a spike-like flare occurs about phase 0, and the source is slightly brighter in the phase  range of 0.3--0.5 \citep{HESS2015}.
The two systems have a few common features in the orbital modulation in X-rays and gamma-rays. 
The TeV light curve is in phase with X-rays, while a different modulation is seen in the GeV band.
As for the variability in the optical and near-infrared bands, \citet{Sarty2011} reported that LS~5039 showed possible orbital modulation with the amplitude of a few mmag in the optical band, while no monitoring in the near-infrared band for gamma-ray binaries hosting an O-type star was reported in the past.

We have been monitoring LS~5039 and 1FGL~J1018.6$-$5856 in the near-infrared ($JHKs$) band as a monitoring campaign of several gamma-ray binaries to search for the orbital modulation associated with the massive star and circumstellar materials \citep{Kawachi2021, Moritani2021}.
The near-infrared flux may change at different orbital phases if the optical depth in the line-of-sight varies significantly across the orbit.
If dust is formed by the wind-wind collision, it may also contribute to the near-infrared flux.

In section 2, we describe the observation and data analysis.
The result is shown in section 3.
In section 4, we compare the upper limit of observed photometric variation and the predicted amplitude of variations due to the interaction between the stellar wind and the compact object in these systems. We also discuss the dust formation rate in section 4.
We conclude this paper in section 5.

\section{Observations and Analysis}

\subsection{Observational Configurations}
The near-infrared observations of LS~5039 and 1FGL~J1018.6$-$5856 were performed in April, May and July 2014 and December 2016, respectively, with the SIRIUS (Simultaneous InfraRed Imager for Unbiased Survey; \cite{Nagayama2003}) on the IRSF (InfraRed Survey Facility) 1.4-m telescope, at the South African Astronomical Observatory in Sutherland, South Africa.
SIRIUS is capable of simultaneous imaging in $JHKs$ bands with a field of view of 7$'$.7 $\times$ 7$'$.7. 

Over the individual observation periods, 84 ($J$), 89 ($H$), and 88 ($Ks$) photometories in 18 nights and 54 ($J$), 45 ($H$), and 53 ($Ks$) photometories in 23 nights were conducted for LS~5039 and 1FGL~J1018.6$-$5856, respectively.
The typical seeing was $\sim 1''.8$ (4 pixels), but there were less-photometric data (e.g., data with bad seeing).
Analysis included all the data regardless of the image quality. 
In order to reduce the variance due to the pixel sensitivity, we applied a dithering method by allocating the targets onto the same position in the detector.

\subsection{Data Reduction and Light Curve Analysis}
The primary data reduction was carried out with a standard pipeline software for the SIRIUS\footnote{https://sourceforge.net/projects/irsfsoftware/}, including non-linearity correction, dark subtraction, and flat fielding.
Aperture photometry was applied using {\tt SIRPHOT}, a photometry tool of the pipeline software based on PyRAF.
Here, the aperture of each frame was adjusted to the FWHM of stars in the FoV. 

Brightness of the target at each observation is calculated using the multiple comparison stars.
The comparison stars were selected among the stars in the same FoV which are labeled with \lq AAA' flags ($S/N \ge 10$) in the 2MASS catalogue, whose colors are standard as main sequence stars.
Setting an observation with stable observing condition as the reference, the relative magnitude to the reference, i.e. the differential magnitude, is obtained for each observation to make the light curve.
The average of the magnitude difference of the comparison stars is subtracted in order to remove a noise from the instrument instabilities.
For the same reason not only the statistical error of the target, but also the standard deviation of the comparison stars are included to calculate the error of each observation
 [see \citet{Moritani2021} for details].

\section{Results}


\begin{figure}
\begin{center}
\FigureFile(90,50){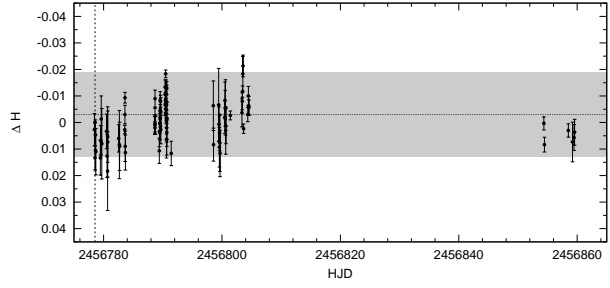}
\caption{Differential magnitude of LS~5039 in $H$ band as a function of HJD. 
The observation on HJD=2456778.55814 is used as the reference of the magnitude (vertical dashed line).
The horizontal dotted line indicates the average magnitude ($-0.003$ mag),
while the range of $2\sigma$ deviation from the average} is marked with shaded area.
\label{fig:LS5039hjd}
\end{center}
\end{figure}


\begin{figure}
\begin{center}
\FigureFile(90,120){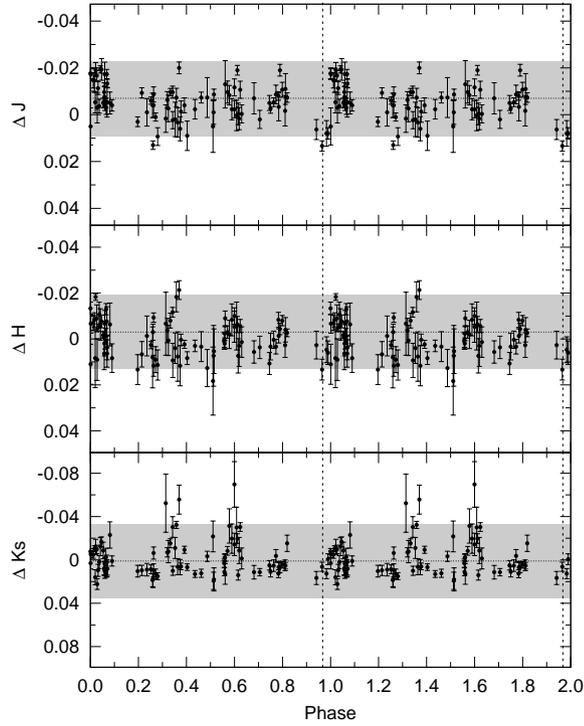}
\caption{Differential magnitude of LS~5039 as a function of orbital phase, in the $J$, $H$, and $Ks$ band from top to bottom. 
The phase of the reference observation corresponds to 0.96727 (vertical dashed line).
The horizontal dotted lines indicate the average magnitude. 
A shaded area is used to indicate the range of $2 \sigma$ deviation from the average.}
\label{fig:LS5039phase}
\end{center}
\end{figure}


\begin{figure}
\begin{center}
\FigureFile(90,50){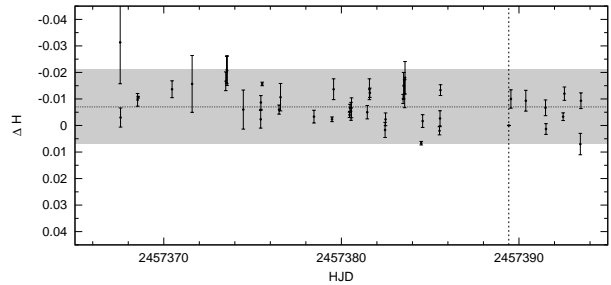}
\caption{The same as Figure \ref{fig:LS5039hjd}, but for 1FGL~J1018.6$-$5856. 
The observation on HJD=2457389.42547 is the reference.
The average of the differential magnitudes is $-0.008$ mag.}
\label{fig:1FGLhjd}
\end{center}
\end{figure}
	

\begin{figure}
\begin{center}
\FigureFile(90,120){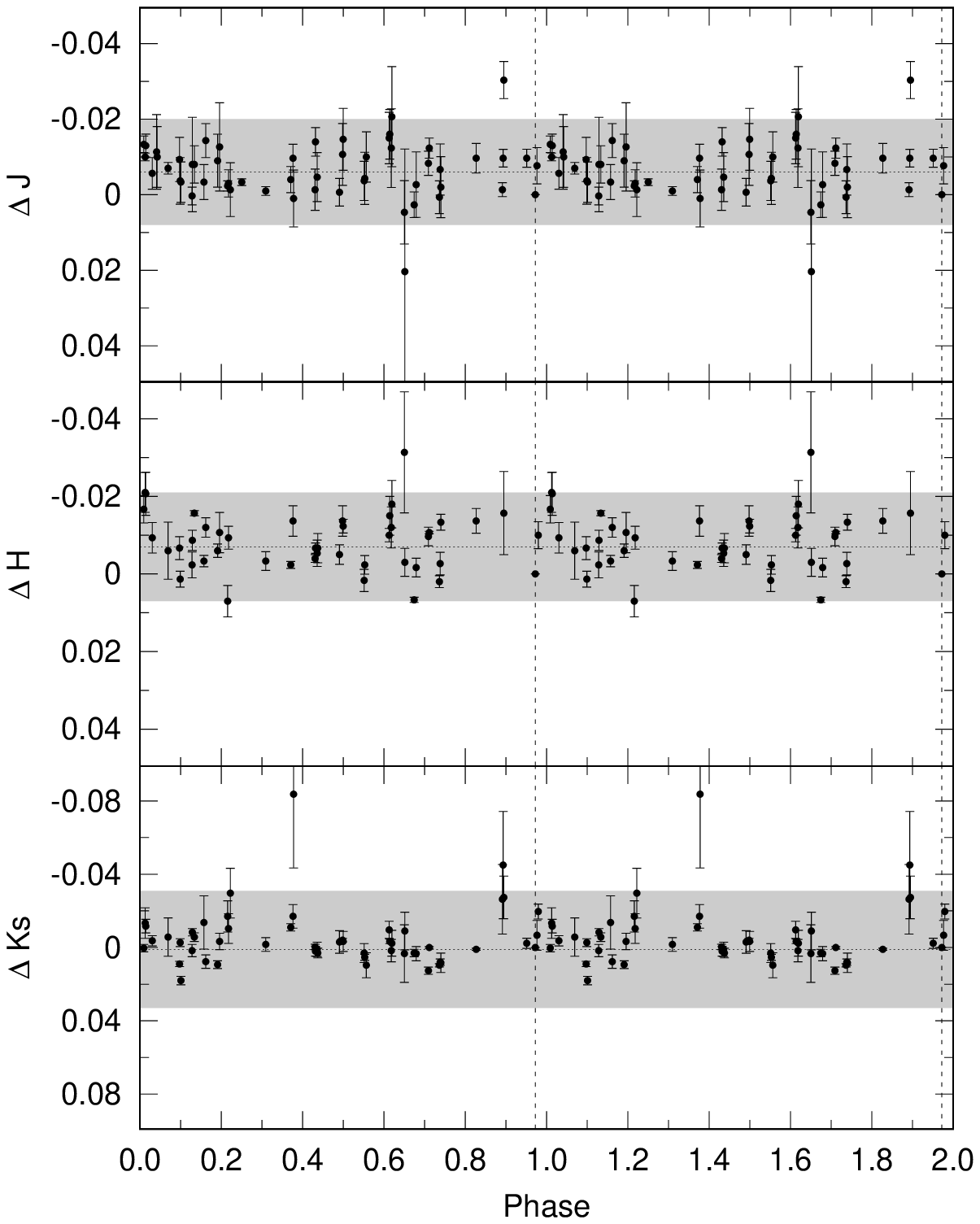}
\caption{The same as Figure \ref{fig:LS5039phase}, but for 1FGL~J1018.6$-$5856.
The phase of the reference observation is 0.9723.
}
\label{fig:1FGLphase}
\end{center}
\end{figure}

Figures \ref{fig:LS5039hjd} and \ref{fig:1FGLhjd} show the $H$-band light curves of LS~5039 and 1FGL~J1018.6$-$5856, respectively, whereas the orbital phasograms of the differential magnitude in the $JHKs$ bands for LS~5039 and 1FGL~J1018.6$-$5856 are respectively shown in Figures \ref{fig:LS5039phase} and \ref{fig:1FGLphase}.
Note that our observations cover the entire orbital phase for both systems.
In these figures, the ordinate shows the differential magnitude, i.e., the magnitude relative to the reference data\, in the observational period.
The average values of the differential magnitudes, which are indicated by the horizontal dotted lines, are ($J$, $H$, $Ks$) = (0.007, $-$0.003, 0.001) mag for LS~5039 and ($-$0.006, $-$0.007, 0.001) mag for 1FGL~J1018.6$-$5856.

As shown in these figures, no clear orbital modulation is detected in the near-infrared band.
There is no significant cycle-to-cycle variation detected nor longer-term ($\sim$ 100 days) variation, either, although observations of 1FGL~J1018.6$-$5856 did not cover many orbital cycles.
We searched LS~5039 data for periodicity other than the orbital period using the Fourier method and the Lomb-Scargle method, but resultant power spectrum had no significant peak.
Furthermore, we conducted a {\textit t}test with a hypothesis that the observed flux is constant at the average value. Here, we subtracted the average value from the individual data.  
The hypothesis is found to be true at a significance level of 5\,\% .

Having said that, it is important to put a constraint on the NIR variability.
If we take the upper limit of the brightness variation to be twice the standard deviation, $\sigma$, the upper limit in the ($J$, $H$, $Ks$) band is obtained as (0.016, 0.016, 0.034) mag for LS~5039 and (0.015, 0.015, 0.031) mag for 1FGL~J1018.6$-$5856. 
It is noted that the variations whose amplitude are smaller than this upper limit of 2$\sigma$ (i.e., 95 \% of the data distribution) might be overlooked in this work. For instance, there might be a hint of variation as in the optical band, although the significance is not sufficient compared to the error. 
Improvement of the measurement error as well as the data coverage through the entire orbital phase are required to discuss such smaller variations.

\section{Discussion}

As shown in the previous section, no significant orbital modulation or variability are detected in the near-infrared photometry data of LS~5039 and 1FGL~J1018.6$-$5856.
Emissions in the optical and near-infrared bands are dominated by thermal emission from the O-star.
Orbital modulations of the optical light curve in HMXBs with normal stars are in general expected to be double-waved because of the tidal distortion. 
In systems with a B star with a circumstellar disk, an optical brightening is observed, which is caused by tidal interaction around the periastron passage \citep{Kawachi2021}.
Figure 4 of \citet{Sarty2011} showed a possible variation in LS~5039 at the level of a few mmag in the optical band (350-750 nm), with an apparent broad minimum at the orbital phase of $\sim0.7-0.8$.
Using the Wilson-Devinney binary star modeling code \citep{Wilson-Devinney1971}, they found that the modeled light curves reasonably represent the broad minimum.
Their best fit parameters ($M_\mathrm{O}=26 ~ \mathrm{M_{sun}}$, $M_\mathrm{co}=1.84 ~ \mathrm{M_{sun}}$, $i=60~^{\circ}$, $e=0.24$) seem consistent with the recent work using $Suzaku$ and $NuSTAR$ data \citep[$M_\mathrm{co}=1.23-2.35 ~ \mathrm{M_{sun}} $, ~ $e=0.278-0.306$]{Yoneda2020}.
If the tidal deformation of the O star causes the optical modulation, a synchronized modulation should also be seen in near-infrared if the photometric accuracy is achieved at the same level (a few mmag or better).

Although we did not detect the orbital modulation in the near-infrared bands, the obtained upper limit of the photometric variability can be utilized to estimate the attenuation of the stellar flux in the stellar wind.
Similar attempt was done by \citet{Sugawara2015} for a Wolf-Rayet+O binary using X-ray flux variation.
It can also be used to constrain the dust formation rate in each system. 
These constraints will provide useful information about these O-star gamma-ray binaries. 
In the following sections we discuss these issues.

\subsection{Near-infrared variability due to wind opacity}
If the wind column density along the line of sight towards the star varies with orbital phase, it might cause near-infrared variability. In order to test this idea, we assume that both LS~5039 and \onefgl\ host a pulsar with a relativistic wind and consider the collision between the stellar and pulsar winds. When the binary is observed at the inferior conjunction at a high inclination angle, the stellar wind is limited inside the radius of the collision shock, whereas it extends to larger distances when observed at other orbital phases. This would result in a lower wind absorption of the stellar flux around the inferior conjunction than at other orbital phases. 
In the following, we study whether this difference is detectable in near-infrared.

We model the stellar wind absorption when the system is viewed edge-on. 
Regarding the inclination angle of LS~5039, the lack of X-ray eclipses suggests $i \lesssim 70^\circ$ (\cite{Reig2003}; see also \cite{Casares2005}). We note that this could be a reasonable value for $i$, but we simplify the model, adopting $i=90^\circ$.
For \onefgl, the recently improved mass function $f(M) = 0.00432\,M_\odot$ \citep{vanSoelen2022} suggests an inclination angle close to $90^\circ$. For instance, we obtain $i=77^\circ$ for $M_* = 22.9\,M_\odot$ and $M_\mathrm{pulsar} = 1.4\,M_\odot$, although lower values of $M_*$ and/or higher values of $M_\mathrm{pulsar}$ allow lower inclination angles.
It is also noted that for a given orbital solution, the mass of a compact object decreases as the inclination angle increases, thus large inclination angles favor a neutron star (pulsar wind model) rather than a black hole (microquasar model) as the nature of the compact object.

For simplicity, we also assume that the stellar wind is intrinsically spherically symmetric,
where the wind mass-loss rate, $\dot{M}$, is connected to the density, $\rho$, and the velocity, $v$, as
\begin{equation}
\dot{M} = 4 \pi r^2 \rho v.
\label{Mdot}
\end{equation}
In what follows, we assume that the wind is isothermal at the stellar effective temperature, 
and has a velocity distribution of the form,
\begin{equation}
v (r) = c_\mathrm{s} + (v_\infty - c_\mathrm{s}) \left(1-\frac{R_*}{r} \right)^\beta,
\label{eq:beta_vel}
\end{equation}
where $c_\mathrm{s}$ is the isothermal sound speed, $v_\infty$ is the wind terminal velocity,
$R_*$ is the radius of the star, and $\beta$ is a parameter that characterizes the acceleration of the wind. 
We adopt a standard value of $\beta = 0.8$. {In our model, we ignore the effect of the Coriolis force on the wind structure.


\begin{figure}
\begin{center}
\includegraphics[width=7cm]{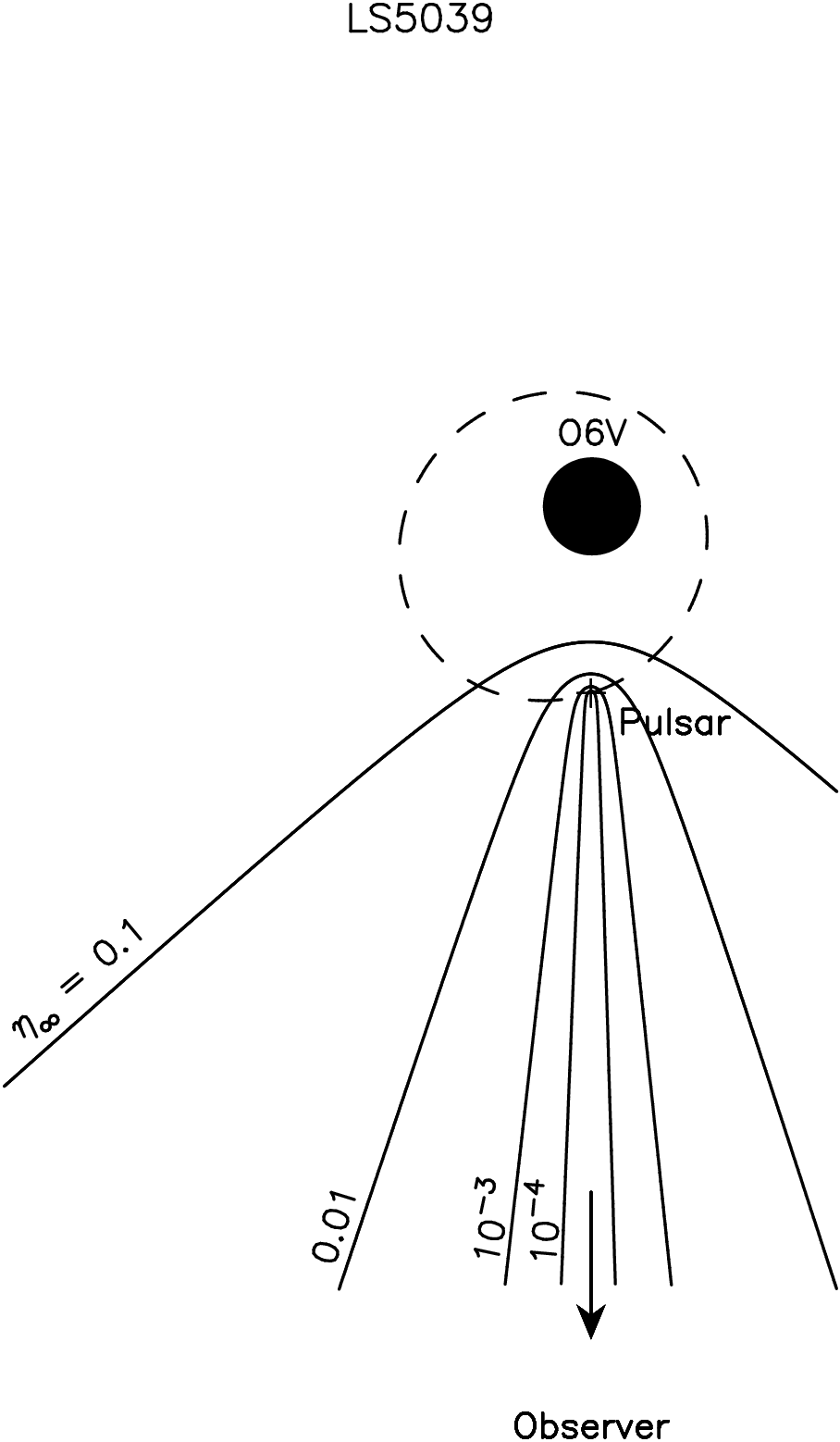}
\hspace*{2em}
\includegraphics[width=7cm]{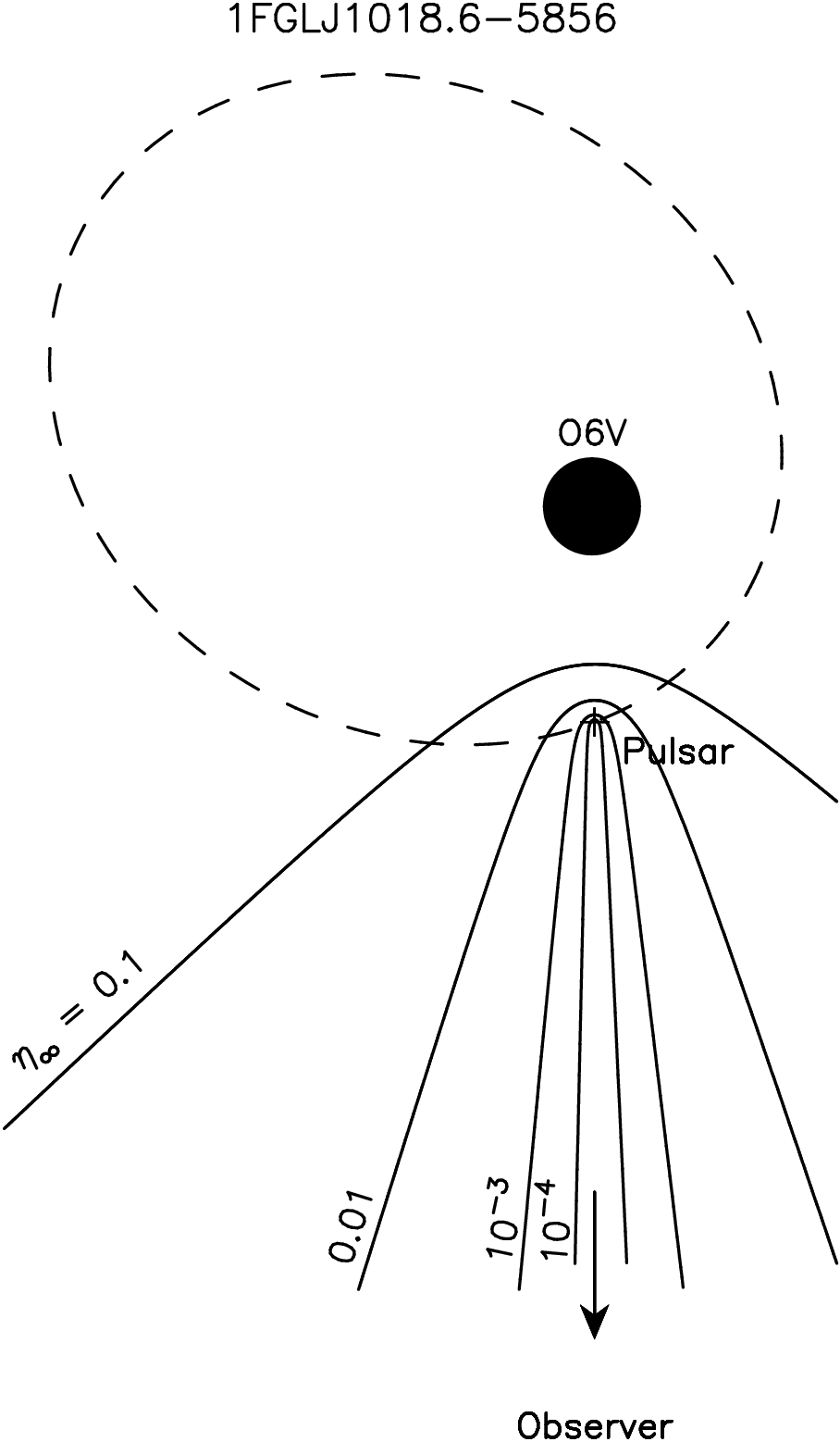}
\caption{Location of the interaction surface between the stellar and pulsar winds for different values of $\eta_\infty \mathbf{(= \dot{E}_\mathrm{pw}/(\dot{M} v_\infty c))}$, overlaid on the orbits of LS~5039 (left) and \onefgl\ (right). The binary configuration is for the inferior conjunction, of which the orbital phase from periastron is 0.716 for LS~5039 \citep{Casares2005} and 0.95 for \onefgl\ \citep{vanSoelen2022}. For comparison purpose, the scale of each panel is normalized by the radius of the O star in each system. Annotated to each interaction surface is the value of $\eta_\infty$, where we adopted $v_\infty = 2,440\,\mathrm{km\,s}^{-1}$ and $2,770\,\mathrm{km\,s}^{-1}$ for LS~5039 and  for \onefgl, respectively. For instance, for $\dot{E}_\mathrm{pw}=10^{36}\,\mathrm{erg\,s}^{-1}$, $\eta_\infty=0.1$, $\ldots$, $10^{-4}$ corresponds to $\dot{M}=2.2 \times 10^{-8}$, $\ldots$, $2.2 \times 10^{-5}\,M_\odot\,\mathrm{yr}^{-1}$ for LS~5039 and $\dot{M}=1.9 \times 10^{-8}$, $\ldots$, $1.9 \times 10^{-5}\,M_\odot\,\mathrm{yr}^{-1}$ for \onefgl, respectively.}
\label{fig:wwc}
\end{center}
\end{figure}

In colliding-wind binaries, if the Coriolis force due to orbital motion is ignored, we have a roughly conical interaction surface across which the momentum fluxes of the two winds balance. On the binary axis, the distance of the interaction surface from the O star, $r_\mathrm{s}$, is obtained from
\begin{equation}
\frac{\dot{M} v(r_\mathrm{s})}{4\pi r_\mathrm{s}^2}
= \frac{\dot{E}_\mathrm{pw}}{4\pi (d-r_\mathrm{s})^2 c}
\label{eq:ram-balance}
\end{equation}
with $\dot{E}_\mathrm{pw}$ and $d$ being the power of the pulsar wind and the binary separation, respectively, as
\begin{equation}
r_\mathrm{s} = \frac{d}{1+\sqrt{\eta(r_\mathrm{s})}},
\label{eq:rs}
\end{equation}
using the momentum ratio, $\eta$, given by
\begin{equation}
\eta(r) = \frac{\dot{E}_\mathrm{pw}}{\dot{M} v(r) c}
     = \frac{\dot{E}_\mathrm{pw}}{\dot{M} \left[ c_\mathrm{s} + (v_\infty - c_\mathrm{s}) \left(1-\frac{R_*}{r} \right)^\beta \right] c}.
\label{eq:eta}
\end{equation}
Figure~\ref{fig:wwc} schematically shows how the location of the interaction surface in LS~5039 and \onefgl\ depends on the relative strength of two winds. Here, we computed the shape of the interaction surface. using the equations in \citet{Antokhin2004}, by solving from the shock apex an ordinary differential equation describing the momentum balance on the surface. For reference, each surface is labeled by the value of $\eta_\infty$, the momentum ratio at infinity, although the location of the surface itself is calculated using the local value of $\eta$. For calculating $\eta_\infty$ in Fig.~\ref{fig:wwc}, we used $v_\infty = 2,440\,\mathrm{km\,s}^{-1}$ \citep{McSwain2004} for LS~5039, whereas for \onefgl, where the wind speed is not known, we took $v_\infty = 2,770\,\mathrm{km\,s}^{-1}$ (three times the escape velocity) as a reasonable value. Note that, in these gamma-ray binaries, $\eta < 1$, so that the interaction surface wraps around the pulsar.

In order to calculate the attenuation of the frequency-dependent stellar flux, we first set up $10^4$ rays parallel to the line of sight in such a way that they are uniformly distributed over the stellar surface that faces the observer. Then, we integrate the wind opacity along each ray from the stellar surface towards the observer, and finally sum up the fluxes over all rays to obtain the total flux to be observed.

As the sources of opacity, we consider electron scattering and free-free absorption.
The optical depth of the wind due to electron scattering, $\tau_\mathrm{es}$, is given by
\begin{eqnarray}
\tau_\mathrm{es} &=& \int \kappa_\mathrm{es} \rho d\ell
   \nonumber\\
   &=& \int \frac{\kappa_\mathrm{es} \dot{M}}
   {4\pi r^2 \left[ c_\mathrm{s} + (v_\infty - c_\mathrm{s}) \left(1-\frac{R_*}{r} \right)^\beta \right]} d\ell,
   \label{eq:tau_es_1}
\end{eqnarray}
where $\kappa_\mathrm{es} = 0.34\,\mathrm{cm}^2\,\mathrm{g}^{-1}$ is the opacity of electron scattering and the integration is done along each ray from its base on the stellar surface through the stellar wind region (see Fig.~\ref{fig:rays}). The integration is terminated when the ray encounters the interaction surface. Otherwise, it is extended to the computational boundary, which is set at $r =5 d_\mathrm{INFC}$, where $d_\mathrm{INFC}$ is the binary separation at the inferior conjunction.
The reason for this choice of computational boundary is as follows.

Although we ignore the effect of the Coriolis force on the structure of the interaction surface for simplicity, recent studies find that due to the Coriolis force, the pulsar wind is terminated in the direction behind the pulsar \citep{Bosch-Ramon2011, Bosch-Ramon2012, Bosch-Ramon2015, Huber2021}. The termination shock occurs at a distance a few times the distance where the lateral ram pressure of the stellar wind, due to the Coriolis acceleration, becomes comparable to the ram pressure of the pulsar wind. The above computational boundary at $r=5d_\mathrm{INFC}$ was taken as a radius smaller than this distance so that bending of the interaction surface due to the Coriolis force can be ignored. Although the structure of the two interacting winds is complicated beyond the termination shock, we assume that the attenuation of the stellar flux due to the stellar wind in this zone is small, given that the absorption coefficient rapidly decreases as $r^{-4}$ [see equations~(\ref{eq:alpha_ff}) and (\ref{eq:tau_ff_2}) below], so most of the absorption occurs close to the star (e.g., for LS~5039 with $\dot{M}=10^{-6.5}\,M_\odot\,\mathrm{yr}^{-1}$, the NIR absorption in $r \le 2R_*$ accounts for 90\% of the total absorption of the case where the stellar wind extends to infinity).


\begin{figure}
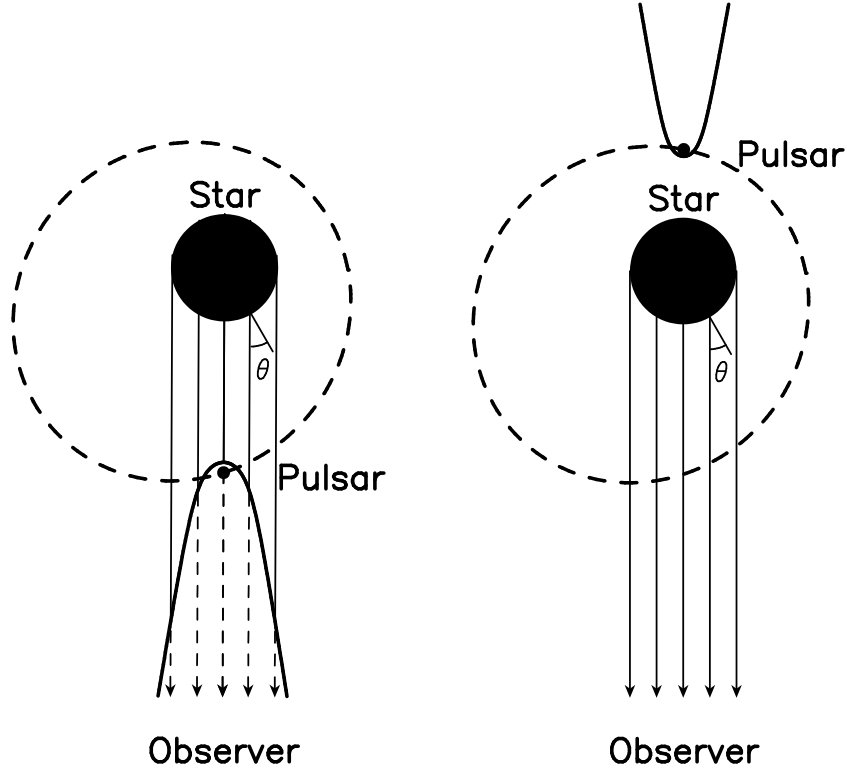

\begin{center}
\includegraphics[width=5cm]{./fig/diagram2_LS5039_infc.eps}
\hspace*{2em}
\includegraphics[width=5cm]{./fig/diagram2_LS5039_supc.eps}
\caption{Schematic diagram showing the location of the interaction surface with respect to the observer.  Left: At the inferior conjunction, the stellar flux is absorbed only in the stellar wind region between the stellar surface and the interaction surface (solid line part of each line of sight), not in the cavity created by the pulsar wind (dashed line part of each line of sight). Right: At the superior conjunction, there is no pulsar wind cavity between the star and the observer, so the stellar flux is absorbed throughout an extended wind region.}
\label{fig:rays}
\end{center}
\end{figure}

The optical depth of the wind due to free-free absorption $\tau_{\mathrm{ff}, \nu}$, is given by
\begin{equation}
\tau_{\mathrm{ff}, \nu} = \int \alpha_{\mathrm{ff}, \nu} d\ell,
\label{eq:tau_ff_1}
\end{equation}
where $\alpha_{\mathrm{ff}, \nu}$ is the free-free absorption coefficient at frequency $\nu$, given by
\begin{equation}
   \alpha_{\mathrm{ff}, \nu} \simeq 3.7 \times 10^{8} T_\mathrm{w}^{-1/2} Z^2 n_\mathrm{e} n_\mathrm{i} 
   \nu^{-3} \left( 1 - e^{-h\nu/kT_\mathrm{w}} \right) \bar{g}_\mathrm{ff}
   \label{eq:alpha_ff}
\end{equation}
in cgs units (cm$^{-1}$) \citep{Rybicki1979}, where $T_\mathrm{w}$ is the temperature of the isothermal wind, $Z$ is the charge of the ions, $n_\mathrm{e}$ and $n_\mathrm{i}$ are respectively the number densities of the electrons and ions, and $\bar{g}_\mathrm{ff}$ is the temperature-averaged Gaunt factor, which is a frequency-insensitive number of order unity.
Adopting $Z^2=1.4$ for an ionized solar composition (e.g., \cite{Dubus2013}), we can rewrite equation~(\ref{eq:tau_ff_1}) as
\begin{eqnarray}
\tau_{\mathrm{ff}, \nu} &\simeq& \mathbf{5.2} \times 10^{8} T_\mathrm{w}^{-1/2} 
\nu^{-3} \left( 1 - e^{-h\nu/kT_\mathrm{w}} \right) \bar{g}_\mathrm{ff}
  \nonumber\\
  && \times \mathbf{\frac{\dot{M}^2}{16\pi^2 \mu_\mathrm{i} \mu_\mathrm{e} m_\mathrm{H}^2}}
  \nonumber\\
  && \times \int \frac{d\ell}
     {r^4 \left[ c_\mathrm{s} + (v_\infty - c_\mathrm{s}) \left(1-\frac{R_*}{r} \right)^\beta 
     \right]^2}.
  \nonumber\\
     \label{eq:tau_ff_2}
\end{eqnarray}
Here, $\mu_\mathrm{i}$ and $\mu_\mathrm{e}$ are the mean ion and electron molecular weights, respectively, and $m_\mathrm{H}$ is the mass of the hydrogen atom.
In the following calculation, we adopt $\mu_\mathrm{i}=1.30$ and $\mu_\mathrm{e}=1.18$ for an ionized solar composition (e.g., \cite{Dubus2013}) and calculate $\bar{g}_\mathrm{ff}$ by a program and data provided by \citet{vanHoof2014}, which yields $\bar{g}_\mathrm{ff}$ in the range of 1.36--1.46 for our model parameters.
As in the electron scattering case, the integration of equation~(\ref{eq:tau_ff_2}) is done along each of $10^4$ rays from the stellar surface to the interaction surface or to the calculation boundary otherwise, depending on whether there is the interaction surface between the star and the observer.

In the wind where both electron scattering and free-free absorption occur, the effective optical depth, $\tau_{\mathrm{eff}, \nu}$, is given by
\begin{equation}
\tau_{\mathrm{eff}, \nu} = \sqrt{\tau_{\mathrm{ff}, \nu} \left[ \tau_{\mathrm{ff}, \nu} + \tau_\mathrm{es} \right]}
\label{eq:tau_effective}
\end{equation}
(e.g., \cite{Rybicki1979}), for which the stellar flux at frequency $\nu$ is attenuated by a factor $\exp (-\tau_{\mathrm{eff}, \nu})$. In the current model where the near-infrared variability is caused by the difference in the wind absorption at the inferior conjunction (INFC) and other phases such as the superior conjunction (SUPC) (see Figure~\ref{fig:rays}), the amplitude of the magnitude variation at frequency $\nu$, $|\Delta m_\nu|$, is obtained by
\begin{equation}
|\Delta m_\nu| = \frac{5}{2}\log_{10} \frac{S_{\nu, \mathrm{INFC}}}{S_{\nu, \mathrm{SUPC}}} = \frac{5}{2}\log_{10}  \frac{\left[\displaystyle \sum_{i=1}^{10^4} \cos \theta_i \exp (-\tau_{\mathrm{eff}, \nu i})\right]_\mathrm{INFC}}{\left[\displaystyle \sum_{i=1}^{10^4} \cos \theta_i \exp (-\tau_{\mathrm{eff}, \nu i})\right]_\mathrm{SUPC}},
\label{eq:delta_m_abs}
\end{equation}
where $S_{\nu, \mathrm{INFC}}$ and $S_{\nu, \mathrm{SUPC}}$ are absorbed stellar fluxes at frequency $\nu$ at INFC and SUPC, respectively. In the last equation, the summation is taken over all rays towards the observer, $i$-th of which has an angle $\theta_i$ from a vector normal to the stellar surface (see Fig.\ref{fig:rays}) and optical depth $\tau_{\mathrm{eff}, \nu i}$ at frequency $\nu$, which are evaluated at INFC and SUPC. Here, we have assumed that the stellar surface has a uniform brightness.

Using equations~(\ref{eq:tau_es_1}), (\ref{eq:tau_ff_2}), and (\ref{eq:tau_effective}), together with the orbit solution and other system parameters for each system (see Table~\ref{tbl:parameters}), to evaluate equation~(\ref{eq:delta_m_abs}), we obtain a relation between the amplitude of the near-infrared modulation and the wind mass-loss rate, under the assumption that the near-infrared variability is dominated by electron scattering and free-free absorption.

\begin{table*}
\caption{System parameters adopted and observed near-infrared variability for LS~5039 and 1FGL~J1018.6$-$5856.}
\label{tbl:parameters}
\begin{center}
\begin{tabular}{ccc}
\hline
\multicolumn{1}{c}{Parameters} & LS~5039 & 1FGL~J1018.6$-$5856 \\
\hline
$M_*$ ($M_\odot$) & 22.9$^1$ & 22.9 (assumed) \\
$R_*$ ($R_\odot$) & 9.3$^1$ & 10.23$^2$ \\
$T_\mathrm{eff}$ (K) & 39,000$^1$ & 38,151$^2$ \\
$T_\mathrm{w}$ (K) & 39,000 (assumed) & 38,151 (assumed) \\
$v_\infty$ ($\mathrm{km\,s}^{-1}$) & 2,440$^3$ & 1,500 -- 3,500 (assumed) \\
$\beta$ & 0.8 (assumed) & 0.8 (assumed) \\
$M_\mathrm{pulsar}$ ($M_\odot$) & 1.4 (assumed) & 1.4 (assumed) \\
$P_\mathrm{orb}$ (d) & 3.096$^1$ & 16.5507$^4$ \\
$e$ & 0.35$^1$ & 0.531$^4$ \\
INFC from periastron & 0.716$^5$ & 0.95$^4$ \\
SUPC from periastron& 0.058$^5$ & 0.155$^4$ \\
$\dot{E}_\mathrm{pw}$ ($\mathrm{erg\,s}^{-1}$) & $10^{36}$ or $10^{37}$ (assumed$^6$) & $10^{36}$ or $10^{37}$ (assumed$^6$) \\
$\Delta J$ (mag) & $<$0.016$^7$ & $<$0.015$^7$ \\
$\Delta H$ (mag) & $<$0.016$^7$ & $<$0.015$^7$ \\
$\Delta Ks$ (mag) & $<$0.034$^7$ & $<$0.031$^7$ \\
\hline
\end{tabular}\\
1~\citet{Casares2005}, 2~Parameters of a normal O6V star taken from \citet{Martins2005}, 3~\citet{McSwain2004}, 4~\citet{vanSoelen2022}, 5~\citet{Aharonian2006}, 6~Results are insensitive to $\dot{E}_\mathrm{pw}$. 7~this work
\end{center}
\end{table*}


\begin{figure}
\begin{center}
\begin{tabular}{cc}
\includegraphics[width=8.cm]{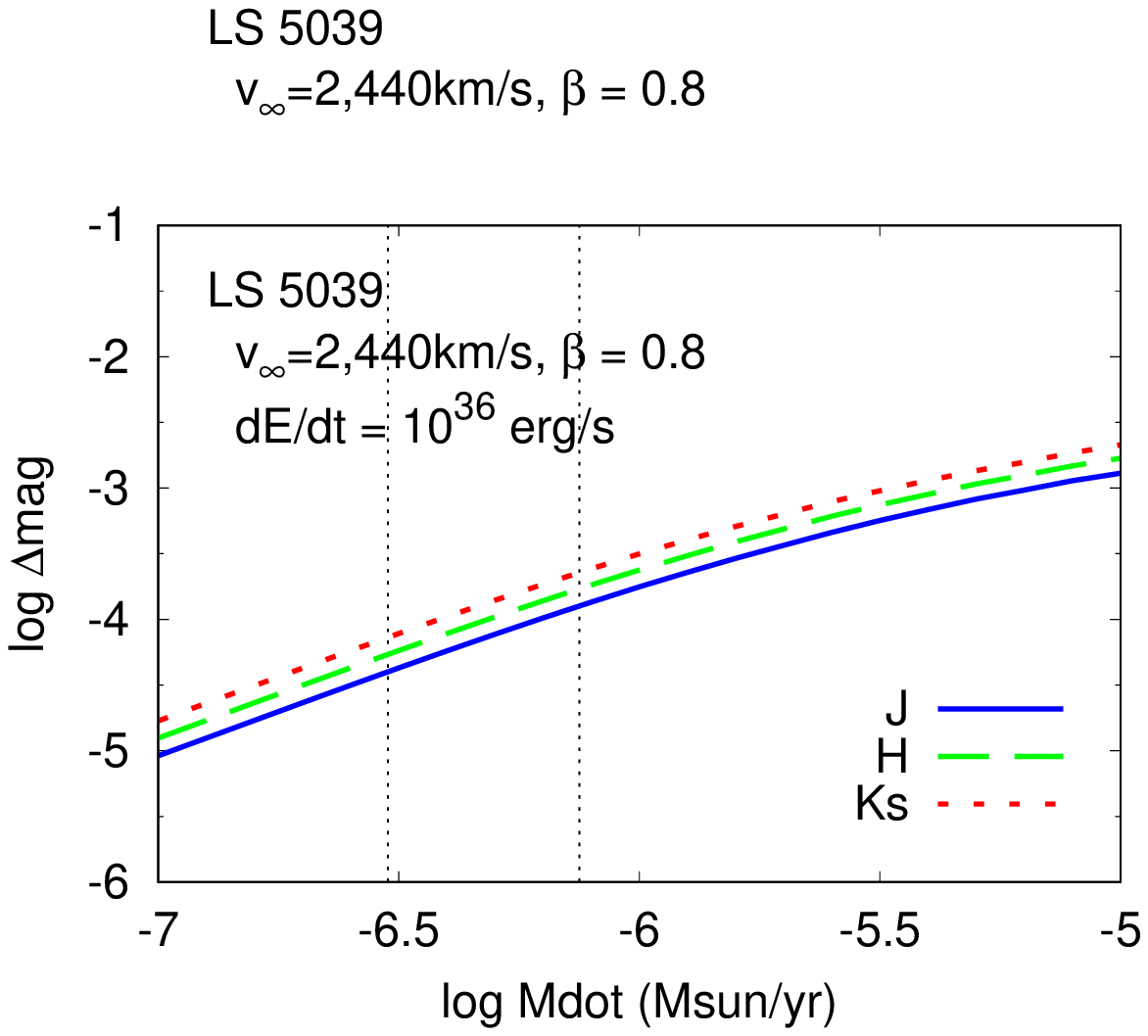} &
\includegraphics[width=8.cm]{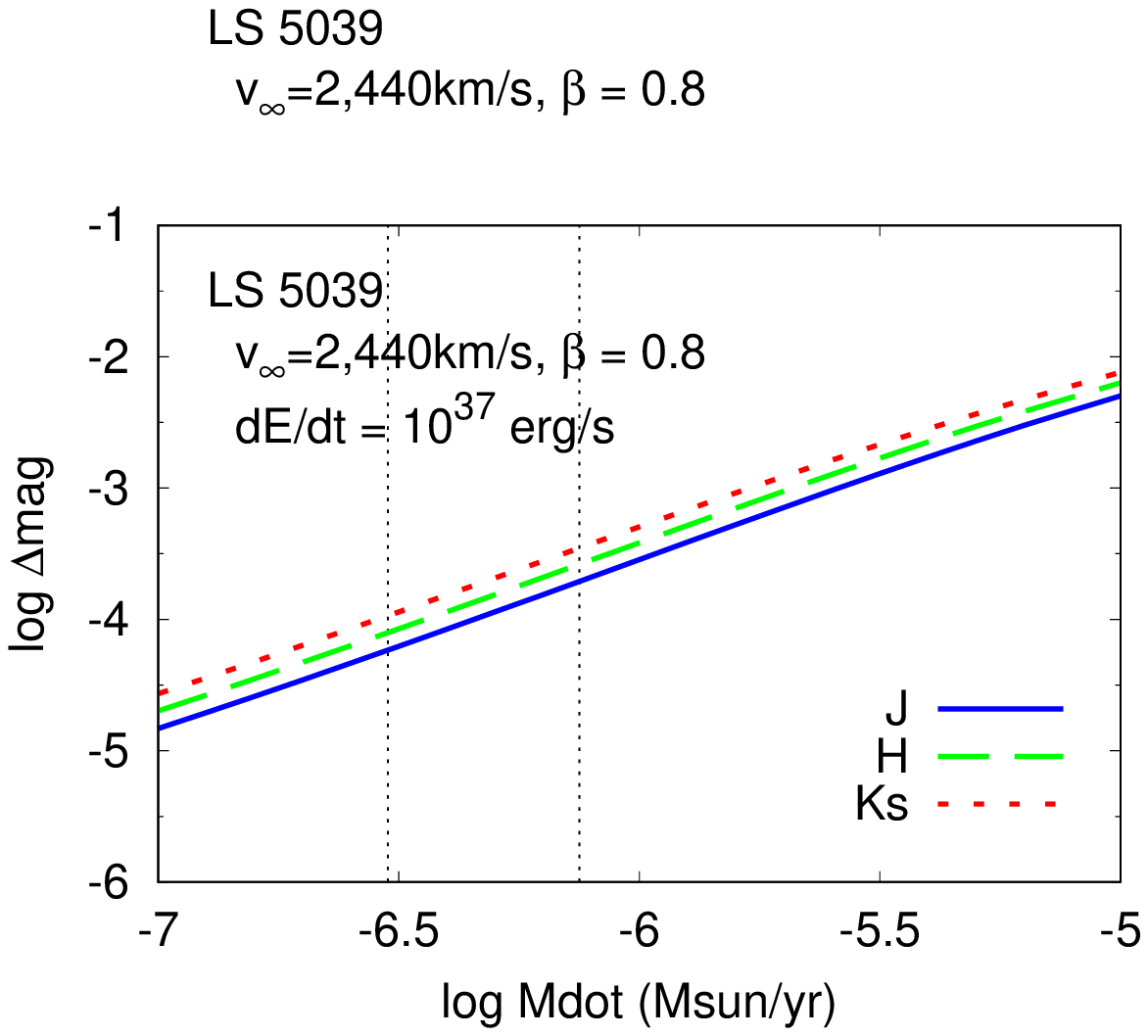}　\\
\includegraphics[width=8.cm]{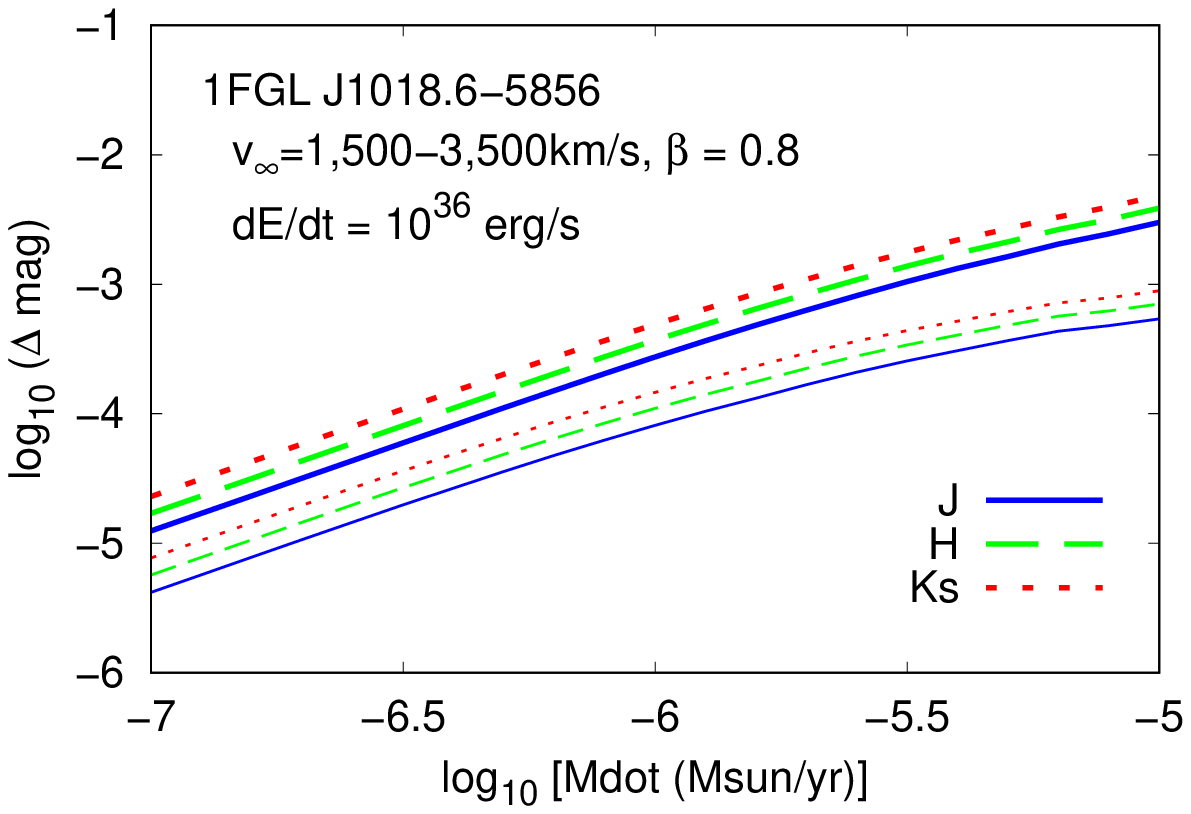} &
\includegraphics[width=8.cm]{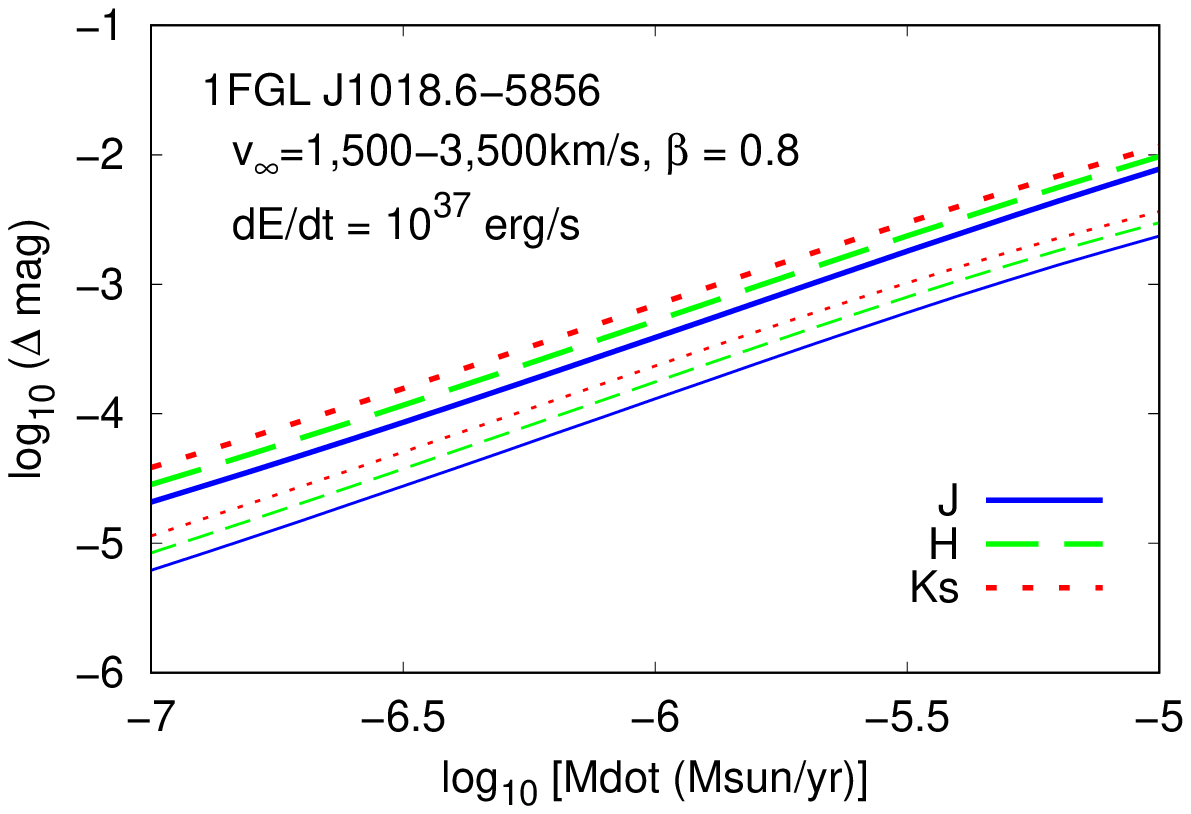}
\end{tabular}
\caption{Mass-loss rate versus near-infrared flux variation caused by electron scattering and free-free absorption in the stellar wind in LS~5039 (upper panels) and \onefgl\ (lower panels). Left panels are for $\dot{E}_\mathrm{pw} = 10^{36}\,\mathrm{erg\,s}^{-1}$ and right panels for $\dot{E}_\mathrm{pw} = 10^{37}\,\mathrm{erg\,s}^{-1}$.
In each panel, the blue solid, green dashed, and red dotted lines are for the $J$, $H$, $Ks$ bands, respectively.
The vertical dotted lines in the upper panels denote the range of wind mass-loss rate, $(3-7.5) \times 10^{-7}\,M_\odot\,\mathrm{yr}^{-1}$, derived from the H$\alpha$ line profiles. In the lower panels for \onefgl, since the terminal velocity has not been determined yet, we plot two cases ($v_\infty = 1,500\,\mathrm{km\,s}^{-1}$ in thick lines and $v_\infty = 3,500\,\mathrm{km\,s}^{-1}$ in thin lines).}
\label{fig:mdot_es_ffa}
\end{center}
\end{figure}

Figure \ref{fig:mdot_es_ffa} shows the amplitude of the near-infrared variability as a function of wind mass-loss rate for LS~5039 (upper panels).
Observationally, the mass-loss rate derived from the H$\alpha$ line profiles is in the range of $(3-7.5) \times 10^{-7}\,M_\odot\,\mathrm{yr}^{-1}$ [$4 \times 10^{-8} - 3 \times 10^{-7}\,M_\odot\,\mathrm{yr}^{-1}$ \citep{McSwain2004}; $(3.7 - 7.5) \times 10^{-7}\,M_\odot\,\mathrm{yr}^{-1}$ \citep{Casares2005}; $3.7 \times 10^{-7}\,M_\odot\,\mathrm{yr}^{-1}$ \citep{Sarty2011}]. For this range of $\dot{M}$, which is marked by two vertical lines in Figure~\ref{fig:mdot_es_ffa}, the amplitudes of variability expected in the $J$, $H$, and $Ks$ bands are respectively $\mathbf{3.9} \times 10^{-5} - \mathbf{1.2} \times 10^{-4}\,\mathrm{mag}$, $\mathbf{5.2} \times 10^{-5} - \mathbf{1.6} \times 10^{-4}\,\mathrm{mag}$, and $\mathbf{6.9} \times 10^{-5} - \mathbf{2.2} \times 10^{-4}\,\mathrm{mag}$ for $\dot{E}_\mathrm{pw}=10^{36}\,\mathrm{erg\,s}^{-1}$, whereas for a stronger $\dot{E}_\mathrm{pw}$ of $10^{37}\,\mathrm{erg\,s}^{-1}$, the photometric variability slightly increases to $\mathbf{5.7} \times 10^{-5} - \mathbf{1.9} \times 10^{-4}\,\mathrm{mag}$, $\mathbf{7.6 \times 10^{-5}} - \mathbf{2.5} \times 10^{-4}\,\mathrm{mag}$, and $(\mathbf{1.0} - \mathbf{3.3}) \times 10^{-4}\,\mathrm{mag}$, respectively. Given that the upper limit of the $J$-, $H$-, and $Ks$-band variability obtained in this work is 0.016~mag, 0.016~mag, and 0.034~mag, respectively, almost two-orders of magnitude better accuracy (i.e., a sub-millimagnitude accuracy) in photometry is needed to detect the wind absorption effect.

The lower panels of Figure \ref{fig:mdot_es_ffa} present the corresponding result for \onefgl. Since the stellar parameters of 1FGL~J1018.6$-$5856 have not been established, we took the stellar radius and temperature from those of a normal O6V star in \citet{Martins2005}.
Since the wind terminal velocity of \onefgl\ is not known either, we studied a range of terminal velocity between $1,500\,\mathrm{km\,s}^{-1}$ and $3,500\,\mathrm{km\,s}^{-1}$, which includes the value $v_\infty \simeq 3 v_\mathrm{esc}$ ($\simeq 2,770\,\mathrm{km s}^{-1}$ in our case) predicted by the radiation-driven wind model \citep{Friend1986, Pauldrach1986}, where $v_\mathrm{esc}$ is the escape velocity given by $v_\mathrm{esc} = \sqrt{2GM_*/R_*}$. The variability amplitudes for the lower and upper ends of this velocity range are shown in each lower panel by the thick and thin lines, respectively. From the figure, it is apparent that the variability decreases with increasing terminal velocity. This is because higher wind speed at a given mass-loss rate means lower wind density and stronger wind momentum flux, both of which works to reduce the variability. Note that for \onefgl, a sub-mmag photometric accuracy is also needed to detect the variability due to wind absorption.

In general, if two colliding wind systems have similar stellar and wind parameters, the one with smaller orbit is likely to show larger photometric modulations, because of the interaction surface being closer to the star. In Figure~\ref{fig:mdot_es_ffa}, however, we note that the relation between the wind mass-loss rate and the amplitude of near-infrared variability are very similar for LS~5039 and \onefgl, despite that as shown in Figure~\ref{fig:wwc}, the orbit of the former is much smaller than that of the latter. The reason is that the amplitude of variability depends on the binary separation at INFC, not on the size of the orbit itself, and that both systems have very similar binary separation at INFC (see Figure~\ref{fig:wwc}).

In summary, wind absorption of the stellar flux in colliding-wind gamma-ray binaries can cause photometric, orbital modulations at a level of 0.1--1~mmag in near-infrared, depending on the mass-loss rate and terminal velocity of the stellar wind. Unfortunately, as of today, the precision in the optical and infrared photometry using ground-based telescope is limited to a few mmag due to atmospheric variation. When observation is done from space, or when careful data processing is done to correct detector characteristics as well as atmospheric variation by fitting several-hour data and/or monitoring water vapor amount, $\sim 0.1-0.2$~mmag precision has been achieved in many cases [e.g., \citet{Clanton2012}]. In general, at the accuracy level of 10--30~mmag, photometric variability due to stellar wind absorption can be seen only in systems where a very strong stellar wind ($\dot{M} \gtrsim 3 \times 10^{-6}\,M_\odot\,\mathrm{yr}^{-1}$ as in a wind of a vary massive O/Wolf-Rayet star) collides with a strong pulsar wind ($\dot{E}_\mathrm{pw} \gtrsim 10^{37}\,\mathrm{erg}\,\mathrm{s}^{-1}$), if there are any such systems.

In addition to the above absorption effect, we have calculated the contribution of free-free emission in the O-star wind to the near-infrared variability, and found that the variability due to free-free emission is more than one order of magnitude smaller than that due to absorption. Since the stellar winds of normal OB stars are optically thin in near-infrared, the free-free emission arises from the whole wind region. In this case the orbital modulation is caused by the variation in the distance and volume of the cavity created by the pulsar wind as the pulsar moves along an eccentric orbit. In our modeling, we adopted the same simplified assumptions as those made for absorption modeling: The computational domain was limited within $r=5d_\mathrm{INFC}$ from the O star, where $d_\mathrm{INFC}$ is the binary separation at the inferior conjunction. Then, for LS~5039, the total luminosity of free free emission from a wind of $\dot{M}=10^{-6}\,M_\odot\,yr^{-1}$ is $\sim 6.3 \times 10^{33}\,\mathrm{erg}\,\mathrm{s}^{-1}$ in the $Ks$ band, which is about 5\% of the stellar luminosity in the same waveband, but its orbital variation amounts to mere 0.05\% of the total emission, resulting in the variability of $4 \times 10^{-6}$ mag. For the $J$ and $H$ bands, the variability is even smaller. For \onefgl, a similar level of variability is obtained.

In evaluating the mass-loss rate of the wind of O stars,  the largest uncertainty is in the effect of small-scale inhomogeneity (clumping). Clumping of O-star winds is expected theoretically (e.g., \cite{Driessen2022}, and references therein). It is also supported observationally. For instance, in HMXBs consisting of supergiants, which show very short ($<1$ day) variabilities in the X-ray light curve, X-ray observations indicate the existence of a dense clumpy wind \citep{Rampy2009,Pradhan2020}.
\citet{Muijres2011} studied the effect of clumping on the stellar wind properties of O-type stars. Assuming that the velocity structure is not affected by clumps, they have found that the mass-loss {rate} decreases unless the clump size {is} very small ($< 1/1000$ of the local density scale height). 
On the other hand, \citet{Sundqvist2019} and \citet{Bjorklund2020} studied the clumping effect by solving full equation of motion locally self-consistently, and found that the mass-loss rate is not significantly affected by clumping, though the terminal velocity increases.
Qualitatively, for a given wind mass-loss rate, clumping will increase the variability caused by free-free absorption, which is sensitive to the square of density. In other words, for a filling factor of $f$, the mass-loss rate is reduced by a factor of $f^{-1/2}$, while keeping wind emission and absorption the same, basically just increasing $\eta_\infty$ by the same factor. Detailed quantitative estimate of that effect, however, is beyond the scope of this paper.

\subsection{Constraints on the dust formation rate}
\label{sec:dust_formation}

Dust grains are strong infrared emitters.
Many Wolf-Rayet binaries show emission from dust formed by the wind-wind collisions (e.g., \cite{Williams2009}). 
Although all Wolf-Rayet binaries showing dust emission host a WC star, which suggest{s} the presence of carbon is crucial (de Becker, private communication), it is too early to rule out the possibility that dust is {also} formed in the O-star wind. 
Indeed \citet{Siebenmorgen2018} reported that a few main-sequence OB stars have high excess in the far-infrared band which suggests existence of warm ($>250 K$) dust close to the star.
They pointed out that dust can survive OB star radiation, if it is formed in a shocked stellar wind region.
\citet{Adams2013} also reported a few dusty OB main sequence stars in SMC.

Even if chances to detect dust emission is low, it is interesting to study possible constraints on the dust formation rate in gamma-ray binaries, from near-infrared observations. 
The following is the first effort along this line of study.

The flux of the dust thermal emission, $F_{\mathrm{d}, \nu}$, is connected to the mass of dust grains, $M_\mathrm{d}$, as 
\begin{equation}
F_{\mathrm{d}, \nu} = \frac{M_\mathrm{d} B_{\nu}(T_\mathrm{d}) \kappa_{\mathrm{d}, \nu}}{D^2}
\label{eq:dust_emission}
\end{equation}
(e.g., \cite{Maeda2013}),
where $B_\nu$ is the Planck function at frequency $\nu$, $T_\mathrm{d}$ is the temperature of dust grains, $\kappa_{\mathrm{d}, \nu}$ is the dust absorption coefficient, which depends on the dust species and size distribution, and $D$ is the distance to the system. 
On the other hand, the stellar flux, $F_{*, \nu}$, is given by
\begin{equation}
F_{*, \nu} = \frac{R_{*}^2 \sigma T_\mathrm{eff}^4}{D^2},
\label{eq:stellar_emission}
\end{equation}
where $\sigma$ is the Stefan-Boltzman constant.

Using equations~(\ref{eq:dust_emission}) and (\ref{eq:stellar_emission}), 
we can relate the modulation in the dust mass $\Delta M_\mathrm{d}$ to the modulation in the photometric magnitude, $\Delta m_\nu$, as follows:
\begin{eqnarray}
\Delta M_\mathrm{d} &=& \frac{R_{*}^2 \sigma T_\mathrm{eff}^4}{B_{\nu}(T_\mathrm{d}) \kappa_{\mathrm{d}, \nu}} \frac{\Delta F_{\mathrm{d}, \nu}}{F_{*, \nu}}
\nonumber\\
&=& \frac{R_{*}^2 \sigma T_\mathrm{eff}^4}{B_{\nu}(T_\mathrm{d}) \kappa_{\mathrm{d}, \nu}} \left( 10^{0.4\Delta m_\nu} - 1 \right).
\label{eq:dustmass_1}
\end{eqnarray}
In the case where no photometric modulation is detected {as in the current study}, the amount of dust formed in one orbital cycle is constrained by changing equation (\ref{eq:dustmass_1}) to an inequality.

The right hand side of equation~(\ref{eq:dustmass_1}) depends on various dust properties, e.g., the dust temperature, the grain species and size distribution.
Since nothing is known about these properties in shock regions in gamma-ray binaries, any estimates based on equation~(\ref{eq:dustmass_1}) inevitably have a large uncertainty. Particularly large uncertainties come from the lack of knowledge on dust temperature and grain species. 
Having said that, we narrow down the possible parameters, with the help of studies available on dust forming Wolf-Rayet binaries. In these systems, the temperature of dust formed at the shock apex is high ($T_\mathrm{d} > 1,000\,\mathrm{K}$) \citep[and references therein]{Marchenko2007}, although there is a system hosting dust in lower temperature \citep{Lau2017}. Moreover, the fact that only WR binaries with a carbon-rich WR star have dust implies that, in the O-star, non carbon-rich wind, astronomical silicate might have more chance to form than does amorphous carbon.

\begin{table*}
\caption{Constraints on the orbital modulation of dust mass, 
$\Delta M_\mathrm{d}$, and the corresponding dust formation rate, 
$\dot{M}_\mathrm{d}$, in the collision shocks in LS~5039 and 1FGL~J1018.6$-$5856.}
\begin{center}
\begin{tabular}{cccccc}
\hline
Waveband & $T_\mathrm{d}$ (K) 
         & \multicolumn{2}{c}{Upper limit of $\Delta M_\mathrm{d}$ ($10^{22}\,\mathrm{g}$)} 
         & \multicolumn{2}{c}{Upper limit of $\dot{M}_\mathrm{d}$ ($10^{-8}\,M_\odot\,\mathrm{yr}^{-1}$)} \\
\cline{3-6}
&& \multicolumn{1}{c}{Carbon} & \multicolumn{1}{c}{Silicate} &
   \multicolumn{1}{c}{Carbon} & \multicolumn{1}{c}{Silicate} \\
\hline
\hline
\multicolumn{6}{c}{LS~5039} \\
\hline
$J$ & 1,000 & 53 & 1,100 & 2.5 & 50 \\
  & 1,500 & 1.1 & 23 & 0.054 &  1.1 \\
\hline
$H$ & 1,000 & 6.4 & 110 & 0.30 & 5.3 \\
  & 1,500 & 0.35 & 6.2 & 0.016 & 0.29 \\
\hline
$Ks$ & 1,000 & 3.3 & 50 & 0.15 & 2.4 \\
   & 1,500 & 0.36 & 5.5 & 0.017 & 0.26 \\
\hline
\multicolumn{6}{c}{1FGL~J1018.6$-$5856} \\
\hline
$J$ & 1,000 & 59 & 1,200 & 0.65 & 13 \\
  & 1,500 & 1.3 & 25 & 0.014 & 0.28 \\
\hline
$H$ & 1,000 & 7.1 & 130 & 0.079 & 1.4 \\
  & 1,500 & 0.38 & 6.8 & 0.0043 & 0.076 \\
\hline
$Ks$ & 1,000 & 3.5 & 54 & 0.039 & 0.60 \\
   & 1,500 & 0.38 & 5.9 & 0.0043 & 0.065 \\
\hline
\end{tabular}
\end{center}
\label{tbl:dust}
\end{table*}

Table~\ref{tbl:dust} presents the constraints on the dust mass modulation and the corresponding dust formation rate in LS~5039 and 1FGL~J1018.6$-$5856, obtained from equation (\ref{eq:dustmass_1}) and the observed upper limit.
The table includes the result for two major grain species, amorphous carbon and astronomical silicate. Since the emission from dust strongly depends on the temperature, and dust emission, if any in collision shocks, is likely dominated by hot dust, we considered only temperatures in the range 1,000-1,500\,K. In Table~\ref{tbl:dust}, the grain size is fixed to 0.01\,$\micron$, because the result depends only weakly on the grain size. 

From Table~\ref{tbl:dust}, we note that the constraint on the dust formation rate is sensitive to both the dust temperature and the observed waveband. This is because the $B_\nu(T_\mathrm{d})$ factor in equation~(\ref{eq:dustmass_1}) varies as $\sim \exp[-h\nu/(kT_\mathrm{d})]$, with $h$ and $k$ being the Planck and Boltzmann constants, respectively, for the near-infrared waveband in the temperature range considered here. As a result, the $Ks$-band observation puts a more stringent constraint on the dust mass modulation and hence the dust formation rate than is done by the $J$- and $H$-band observations. These constraints become stronger with increasing dust temperature. If we adopt $T_\mathrm{d} = 1,500\,\mathrm{K}$, the temperature close to the sublimation temperature, and astronomical silicate as the formed grain species, the upper limit of the dust formation rate, $\dot{M}_\mathrm{d}$, obtained by the current observations is in the order of $\dot{M}_\mathrm{d} \sim 10^{-9} - 10^{-8}\,M_\odot\,\mathrm{yr}^{-1}$. For amorphous carbon, the upper limit is about one order of magnitude lower, i.e., $\dot{M}_\mathrm{d} \sim 10^{-10}-10^{-9}\,M_\odot\,\mathrm{yr}^{-1}$. 
LS~5039 shows {less stringent upper limit on the} dust formation rate than 1FGL~J1018.6$-$5856 because of the shorter orbital period.
In any case, neither binaries are efficient dust makers, compared to $\dot{M}_\mathrm{d} > 10^{-8}\,M_\odot\,\mathrm{yr}^{-1}$ derived for some Wolf-Rayet binaries with dust emission.

\section{Conclusions}
$JHKs$ photometric monitoring was carried out for the gamma-ray binaries LS~5039 and 1FGL~J1018.6$-$5856 using IRSF/SIRIUS searching for orbital modulations.
The observation was performed for more than one orbital period, covering the entire orbital phase densely.
Neither of the two systems showed significant modulation, with the upper limit of $\sim 0.01-0.03$ mag.
Assuming that both systems have a pulsar with a relativistic wind, we calculated the amplitude of photometric variation in near-infrared caused by the difference in the optical depth of O-star wind at the inferior conjunction and the other orbital phases. 
Different from the other orbital phases, at the inferior conjunction the cavity created by the pulsar wind is located between the O star and the observer, so that stellar flux is absorbed by stellar wind only between the stellar surface and the interaction surface.
For a range of the stellar wind mass-loss rate, the amplitude of variation is in sub-mmg range, because in these systems the O-star wind dominates the pulsar wind with a typical power of $\sim 10^{36}\,\mathrm{erg\,s}^{-1}$.
Besides, dust formation rate in colliding wind shock in these gamma-ray binaries was estimated from the upper limit of the photometric variation for the first time. The estimated value of $< 10^{-8}\,M_\odot\,\mathrm{yr}^{-1}$ for the amorphous carbon and the astronomical silicate is much smaller than in WR stars.

\section*{Funding}
This project was supported as a Joint Research Project under agreement between the Japan Society for the Promotion of Science (JSPS) and National Research Foundation (NRF) of South Africa.
This work was also supported by JSPS KAKENHI Grant Numbers JP21540304, JP20540236, JP24540235, and JP21K03619, Research Fellowships for the Promotion of Science for Young Scientists, and a research fund from Hokkai-Gakuen Educational Foundarion.
The IRSF project is a collaboration between Nagoya University and the South African Astronomical Observatory (SAAO) supported by the Grants-in-Aid for Scientific Research on Priority Areas (A) (No. 10147207 and No. 10147214), Optical \& Near-Infrared Astronomy Inter-University Cooperation Program, from the Ministry of Education, Culture, Sports, Science and Technology (MEXT) of Japan, and the NRF of South Africa.

\section*{Acknowledgments}

We thank Takaya Nozawa for kindly providing us data necessary for 
dust emission calculation, including tables of dust absorption coefficient 
for various grain species and sizes.



\begin{thebibliography}{99}
\bibitem[Abdo et al.(2009)]{Abdo2009}
     Abdo, A.~A.., et al., 2009, \apj, 706, 56
\bibitem[Adams et al.(2013)]{Adams2013}
     Adams, J. J., et al., 2013, \apj, 771, 112
\bibitem[Aharonian et al.(2006)]{Aharonian2006}
     Aharonian, F., et al. (HESS Collaboration), 2006, \aap, 460, 743
\bibitem[Antokhin et al.(2004)]{Antokhin2004}
     Antokhin, I.~I., Owocki, S.~P., \& Brown, J.~C., 2004, \apj, 611, 434
\bibitem[Bosch-Ramon et al.(2005)]{Bosch-Ramon2005} 
     Bosch-Ramon, V., Paredes, J.M., Rib\'{o}, M., Miller, J.M., Reig, P., Mart\'{i}, J., 2005, \apjl, 628, 388
\bibitem[Bosch-Ramon \& Barkov(2011)]{Bosch-Ramon2011} 
     Bosch-Ramon, V. \& Barkov, M.~V., 2011, \aap, 535..A20
\bibitem[Bosch-Ramon et al.(2012)]{Bosch-Ramon2012} 
     Bosch-Ramon, V., Barkov, M.~V., Khangulyan, D., \& Perucho, M., 2012, \aap, 544..A59
\bibitem[Bosch-Ramon et al.(2015)]{Bosch-Ramon2015} 
     Bosch-Ramon, V., Barkov, M.~V., Perucho, M., \& Perucho, M., 2015, \aap, 577..A89
\bibitem[Bosch-Ramon \& Barkov(2016)]{Bosch-Ramon2016} 
     Bosch-Ramon, V., Barkov, M. V., 2016, \aap, 590, A119
\bibitem[Bj{\"o}rklund et al.(2020)]{Bjorklund2020} 
     Bj{\"o}rklund, R., Sundqvist, J.~O., Puls, J. \& Najarro, F., 2020, \aap, accepted
\bibitem[Casares et al.(2005)]{Casares2005}
     Casares, J., Rib\'{o} M., Ribas I., Paredes J. M., Mart\'{i} J., Herrero, A., 2005, \mnras, 364, 899
\bibitem[Chen et al.(2021)]{Chen2021} 
     Chen, A.-M., Ng, C., Takata, J. \& Yu, Y.-W., 2021, Research in Astronomy and Astrophysics, 21, 189
\bibitem[Clanton et al.(2012)]{Clanton2012}
     Clanton, C., Beichman, C., Vasisht, G., Smith, R., \& Gaudi, B.~S., 2012, \pasp, 124, 700
\bibitem[Driessen et al.(2022)]{Driessen2022}
     {Driessen, F. A.,  Sundqvist, J. O., \&  Dagore, A., 2022, \aap, 663, A40}
\bibitem[Dubus(2006)]{Dubus2006} 
     Dubus, G., 2006, \aap, 456, 801
\bibitem[Dubus(2013)]{Dubus2013}
     Dubus, G., 2013, \aapr, 21, 64
\bibitem[Friend \& Abbott(1986)]{Friend1986}
     Friend, D. B. \& Abbott, D. C., 1986, \apj, 311, 701
\bibitem[H.~E.~S.~S. Collaboration(2015)]{HESS2015}
     H. E. S. S. Collaboration;, 2015, \aap, 577, 131
\bibitem[Huber et al. (2021)]{Huber2021}
  Huber, D., Kissmann, R., \& Remier, O., 2021, \aa, 649, A71
\bibitem[Kawachi et al.(2021)]{Kawachi2021}
     Kawachi, A. et al. 2021, \pasj, { 73, 545}
\bibitem[Lau et al.(2017)]{Lau2017}
     Lau, R. M. et al. 2017, \apjl, 835, L31
\bibitem[Maeda et al.(2013)]{Maeda2013}
     Maeda, K. et al., 2013, \apj, 776:5
\bibitem[Marcote et al.(2018)]{Marcote2018}
     Marcote, B., Rib\'{o}, M., Paredes, J. M., Mao, M. Y. \& Edwards, P. G., 2018, \aap, 619, 26
\bibitem[Martocchia et al.(2005)]{Martocchia2005}
     Martocchia, A., Motch, C. \& Negueruela, I., 2005, \aap, 430, 245
\bibitem[Marchenko \& Moffat(2007)]{Marchenko2007}
     Marchenko, S. V \& Moffat, A. F. J., 2007, \asp, 367, 213
\bibitem[Martins et al.(2005)]{Martins2005}
     Martins, F. and Schaerer, D. \& Hillier, D.~J., 2005, \aap, 436, 1049
\bibitem[McSwain et al.(2004)]{McSwain2004}
     McSwain, M.V., Gies, D.R., Huang, W., Wiita, P. J., Wingert, D. W. \& Kaper, L., 2004, \apj, 600, 927
\bibitem[Monageng et al.(2017)]{Monageng2017}
     Monageng, I. M., McBride, V. A., Townsend, L. J., Kniazev, A. Y., Mohamed, S. and B{\"o}ttcher, M., 2017, \apj, 847, 68
\bibitem[Moritani and Kawachi (2021)]{Moritani2021}
     Moritani, Y., Kawachi, A., 2021, Universe, 7, 320
\bibitem[Muijres et al.(2011)]{Muijres2011}
     Muijres, L.~E., de Koter, A., Vink, J.~S., Krti\v{c}ka, J., Kub\'{a}t, J. \& Langer, N., 2011, \aap, 526, 32
\bibitem[Nagayama et al.(2003)]{Nagayama2003}
     Nagayama, T. et al., 2003, proc. of SPIE 4841, 459
\bibitem[Paredes \& Bordas(2019)]{Paredes2019}
     Paredes, J.M. \& P. Bordas, 2019, proc. of Frontier Research in Astrophysics - III., Mondello, May 2018, 44
\bibitem[Paredes et al.(2000)]{Paredes2000}
     Paredes, J.M., Mart\'{i}, J., Rib\'{o}, M., \& Massi, M., 2000, Science, 288, 2340
\bibitem[Paredes et al.(2002)]{Paredes2002}
     Paredes, J.M., Rib\'{o}, M., Ros, E., Mart\'{i}, J., \& Massi, M., 2002, \aap, 393, L99
\bibitem[Pauldrach et al.(1986)]{Pauldrach1986}
     Pauldrach, a., Puls, J., \& Kudritzki, R. P., 1986, \aap, 164, 86
\bibitem[Pradhan et al.(2020)]{Pradhan2020}
     Pradhan, P., Bozzo, E. Paul, B., Manousakis, A. \& Ferrigno, C., 2019, \apj, 883, 116
\bibitem[Rampy et al.(2009)]{Rampy2009}
     Rampy, R.~A., Smith, D.~M. \& Negueruela, I., 2009, \apj, 707, 243
\bibitem[Reig et al.(2003)]{Reig2003}
     Reig1, P., Rib\'{o}, M., Paredes, J.~M., and Mart\'{i}, J., 2003, \aap, 405, 285
\bibitem[Rib\'{o} et al.(2002)]{Ribo2002}
     Rib\'{o}, M., Paredes, J.~M., Romero, G.~E., Benaglia, P., Mart\'{i}, J., Fors, O. \& Garc\'{i}a-S\'{a}nchez, J.,  2002, \aap, 384, 954
\bibitem[Rybicki \& Lightman(1979)]{Rybicki1979}
     Rybicki, G. B. \& Lightman, A. P., 1979, \emph{Radiation Processes in Astrophysics}, John Wiley \& Sons, Inc., , p. 162
\bibitem[Sarty et al.(2011)]{Sarty2011}
     Sarty, G. E. et al., 2011, \mnras, 411, 1293
\bibitem[Siebenmorgen el al.(2018)]{Siebenmorgen2018}
     Siebenmorgen, R. Scicluna, P. and Kre{\l}owski, J., 2018, \aap, 620, 32 
\bibitem[Sierpowska-Bartosik \& Torres(2007)]{Sierpowska-Bartosik2007}
     Sierpowska-Bartosik A. \& Torres D.F., 2007, \apjl, 671, 145
\bibitem[Sugawara et al.(2015)]{Sugawara2015}
     Sugawara, Y. et al., 2015 \pasj, 67, 121
\bibitem[Sundqvist et al.(2019)]{Sundqvist2019}
     Sundqvist, J.~O., Bj{\"o}rklund, R., Puls, J. \& Najarro, F., 2019, \aap, 632,126
\bibitem[Takahashi et al.(2009)]{Takahashi2009}
    Takahashi, T. et al. 2009, \apj, 697, 592
\bibitem[van~Hoof et al.(2014)]{vanHoof2014}
     van Hoof, P.A.M., Williams, R.J.R., Volk, K., Chatzikos, M., Ferland, G.J., 
     Lykins, M., Porter, R.L., Wang Y., 2014, \mnras, 444, 420
\bibitem[van~Soelen et al.(2022)]{vanSoelen2022}
     van Soelen, B., McKeague, S., Malyshev, D., Chernyakova, M., Komin, N., Matchett, N., Monageng, I. M.,
     2022, \mnras, 515, 1078
\bibitem[Volkov et al.(2021)]{Volkov2021}     
     Volkov, I., Kargaltsev, O., Younes, G. Hare, J., Pavlov, G., 2021, \apj, 915, 61
\bibitem[Williams et al.(2009)]{Williams2009}
     Williams, P.~M et al., 2009, \mnras, 395, 1749
\bibitem[Wilson \& Devinney(1971)]{Wilson-Devinney1971}
     {Wilson, R. E. \& Devinney, E. J., 1971, \apj, 166, 605}
\bibitem[Yoneda et al.(2020)]{Yoneda2020}
     Yoneda, H., Makishima, K., Enoto, T., Khangulyan, D., Matsumoto, T. and Takahashi, T., 2020, \prl, 125, 111103
\end{thebibliography}
\end{document}